\documentclass[aps,twocolumn,groupedaddress,showpacs]{revtex4}
\usepackage{epsfig,amsbsy,bm,amssymb}
\usepackage{hyperref}

\newcommand{\hs}{\hspace*}
\newcommand{\vs}{\vspace*}
\newcommand{\np}{\newpage}
\newcommand{\cx}{C$_{60}$~}
\newcommand{\eref}[1] {(\ref{#1})}

\newcommand{\Fref}[1] {Fig. \ref{#1}}
\newcommand{\ra}{\rangle}
\newcommand{\la}{\langle}
\newcommand{\nn}{\nonumber}
\newcommand{\be}{\begin{equation}}
\newcommand{\ee}{\end{equation}}
\newcommand{\br}{\begin{eqnarray*}}
\newcommand{\er}{\end{eqnarray*}}
\newcommand{\ba}{\begin{eqnarray}}
\newcommand{\ea}{\end{eqnarray}}
\newcommand{\bp}{\begin{minipage}}
\newcommand{\ep}{\end{minipage}}

\newcommand{\isum}%
{\mathop{\hbox{$\displaystyle\sum\kern-13.2pt\int\kern1.5pt$}}}

\begin{document}
\bibliographystyle{apsrev}


\title {Attosecond time delay in the photoionization of endohedral atoms
A@C$_{60}$: \\ A new probe of confinement resonances}

\author{
P. C. Deshmukh$^{\dagger\ddagger}$, A. Mandal$^\dagger$, S. Saha$^\dagger$,
A. S. Kheifets$^\star$, V. K. Dolmatov$^\flat$ and S T Manson$^\sharp$
}

\affiliation{$^\dagger$Department of Physics, Indian Institute of Technology
  Madras, Chennai-600036, India}

\affiliation{$^\ddagger$Department of Physics, CPGS, Jain University, Bangalore,
560011, India}

\affiliation{$^\star$Research School of Physics and Engineering,
The Australian National University,
Canberra ACT 0200, Australia}

\affiliation{$^\flat$  Department of Physics and Earth Science, University of
  North Alabama, Florence, AL 35632, USA}

\affiliation{$^\sharp$Department of Physics and Astronomy, Georgia State
  University, Atlanta, GA 30303, USA }

\date{\today}
\begin{abstract}
The effects of confinement resonances on photoelectron group delay
(Wigner time delay) following ionization of an atom encapsulated
inside a C$_{60}$ cage have been studied theoretically using both
relativistic and non-relativistic random phase approximations.  The
results indicate clearly the resonant character of the confinement
oscillations in time delay of the $4d$ shell of Xe@\cx and  present a
most direct manifestation of Wigner time delay. These oscillations
were missed in a previous theoretical investigation of Ar@\cx [PRL
{\bf 111}, 203003 (2013)].
\end{abstract}

\maketitle

Unprecedented advances in experimental techniques in measuring time
 intervals at the attosecond level \cite{Baltuska2003} have engendered
 the ability to scrutinize the time delay in photoionization of atomic
 systems in the laboratory
 \cite{P.Eckle12052008,M.Schultze06252010,PhysRevLett.106.143002},
 thereby allowing us to probe the fundamental process of
 photoionization in the time domain.  Specifically, using attosecond
 pulses of electromagnetic radiation, the time difference between the
 emergence of photoelectrons from two neighboring atomic subshells has
 been measured both in Ne \cite{M.Schultze06252010} and Ar
 \cite{PhysRevLett.106.143002,PhysRevA.85.053424}.  These experimental
 results have stimulated a host of theoretical calculations to explain
 and to further explore this phenomenon \cite{PhysRevLett.105.073001,
 PhysRevLett.105.233002,PhysRevLett.107.213605,
 PhysRevLett.108.163001a}.  This is of great interest, not only as a
 new way to study a fundamental process of nature, but also as an
 outstanding, unique opportunity towards a deeper understanding of the
 most informative parameter of the process, the photoionization
 amplitude. This is because the time delay is related to the energy
 derivative of the phase of the amplitude driving the process
 \cite{deCarvalho200283}. Indeed, to date, the only method for getting
 the maximum experimental information on photoionization lies through
 a set of measurements of total and differential photoionization cross
 sections, but allows only the absolute values and relative phases of
 matrix elements to be deduced; this is known as a \textit{complete
 photoionization experiment} \cite{0953-4075-37-6-010}. Time delay
 investigations, however, go beyond the \textit{complete experiment}
 strategy and yield the derivative of the phase with respect to the
 photoelectron energy. Time delay investigations, thus, provide a new
 avenue to discern the characteristics of the basic physical quantity
 - the photoionization amplitude - and, thus, of the photoionization
 phenomenon itself. It is the ultimate aim of this paper to promote
 the expansion of time delay studies towards situations where they
 have not yet been exploited and where novel effects might occur - to
 atoms under confinement.

The theory of time delay in physics was developed some time ago
\cite{PhysRev.98.145} and was originally envisioned as a way to study
resonances - the temporary trapping of one (or more) electrons in a
quasi-bound state or a potential well.  Indeed, the Breit-Wigner
formula of resonant scattering $\tau=2/\Gamma$ equates the time delay
$\tau$ with the resonant width $\Gamma$ at half maximum of the
cross-section \cite{Newton1982}.
Resonances are ubiquitous in photoionization of atoms, and these
resonances can be of different natures: inner-shell excitations,
two-electron excitations, shape resonances, etc.  Of great recent 
interest to investigators, which has shaped an area of extremely
active modern research activities \cite{Saunders05031993}, have been
studies of a new genre of resonances, termed confinement resonances, that
occur in the photoionization of an atom $A$ trapped at the center of a
\cx  molecule, the $A@C_{60}$ endohedral atom. The phenomenon of
confinement resonances was predicted theoretically long ago (see,
e.g., \cite{PhysRevA.47.1181}).  However, it is only fairly recently
that confinement resonances have been studied in depth in numerous
theoretical studies at various levels of sophistication (see 
\cite{0953-4075-38-10-L06,0953-4075-41-16-165001,
Dolmatov2009,PhysRevA.81.013202} and references therein), and only
very recently verified experimentally in the photoionization of
endohedral atoms \cite{PhysRevLett.105.213001,PhysRevA.88.053402}. Confinement resonances
have been explained as interferences between the photoelectron wave
emitted directly, and those that experience one or more scattering off
the walls of the encapsulating fullerene \cite{Luberek1996147}. 

If this understanding is correct, these multiple scatterings should
show up prominently in the time delay of the photoelectron relative to
the time delay of the free atom. It is, however, not at all clear
beforehand as to what degree time delay of atomic photoionization is
modified by confining an atom inside of \cx compared to the free atom.
Moreover, a recent theoretical study of time delay in Ar@\cx
\cite{PhysRevLett.111.203003} have not revealed any confinement
resonances in the time delay at all. At the same time, confinement
resonances were clearly seen in time delay of a model system
He$^+$@\cx \cite{Nagele2014}. This system, however, was unstable and
the theoretical analysis was performed on a one-electron level.  We
are, therefore, presented with the task of unraveling the confinement
resonances in a realistic atomic system taking full account of
many-electron effects, thereby significantly enlarging
upon the previous studies \cite{PhysRevLett.111.203003,Nagele2014}.

In this Letter, we explore photoionization of the $4d$ subshell of
Xe@\cx, where confinement resonances have been found experimentally
\cite{PhysRevLett.105.213001,PhysRevA.88.053402}.  Specifically, we
focus upon how the time delay can be used to characterize the
confinement resonances, along with the time delay phenomenology
produced by the resonances.  A model potential is employed to
introduce the effects of the confining \cx on the encaged Xe atom.  In
this model, the C$_{60}$ cage is approximated by an attractive
spherical square well potential of certain inner radius $R_{\rm
inner}$, thickness $\Delta$, and depth $U_{0}$:
\be
V(r)=\left\{
\begin{array}{rl}
-U_{0} <0 & {\rm if~} R_{\rm inner} \le r \le R_{\rm inner}+\Delta
\\
0 & {\rm otherwise.} 
\end{array}
 \right. 
\label{eqVc}
\ee
This should be adequate because the $4d$
subshell is so deeply bound that it cannot hybridize with any of the
levels of \cx, and the photon energy range (80 to 160~eV) is well away
from the \cx plasmons so that interchannel coupling with atomic
photoionization is not important.  Furthermore, recent photoionization calculations
employing such a model resulted in rather good agreement with the
experimental confinement resonances  \cite{PhysRevA.88.053402}.  To
properly account for correlations, the calculations are performed
within the framework of the random-phase approximation (RPA), both the
non-relativistic \cite{AC97} and relativistic
\cite{PhysRevA.20.964} versions. This has the advantages of
spotlighting any relativistic effects.

\begin{figure}[htbp] \centering
  \includegraphics[width=8.0cm] {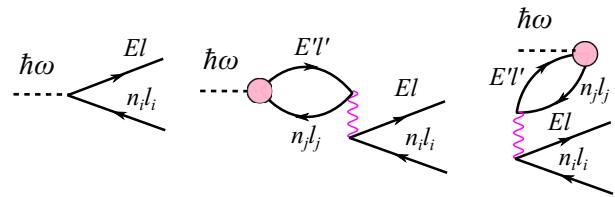} 
  \caption{(Color online) Graphical representation of the RPA
    Equations \cite{AC97}.  Left: non-correlated dipole matrix element.
    Center: time-forward process.  Right: time-reverse process.}
\label{Feynman}
\end{figure}

The time delay is calculated from the photoionization amplitude as 
$
\tau = d\,{\arg f(E)}/dE\equiv
{\rm Im} \Big[ f'(E)/f(E) \Big] \ .
$
The  amplitude $f(E)$ is given by the partial wave
expansion as
\ba
\label{amplitude}
f(E)&\propto&
\sum_{l=l_i\pm1}
e^{i\delta_l}i^{-l}
Y_{lm}(\hat  k)\,
(-1)^m
\left(\begin{array}{rrr}
l&1&l_i\\
-m&0&m_i\\
\end{array}\right)
\nn\\&&\hs{2cm}\times \
 \la El\|D\|n_il_i \ra
\ea
evaluated in the $\hat z$ direction of the polarization axis of light.
In the non-relativistic RPA method, the reduced dipole matrix element
$\la El\|D\|n_il_i \ra$ is evaluated by solving a set of
integro-differential equations \cite{AC97} exhibited graphically by
diagrams of \Fref{Feynman}. In the absence of inter-shell correlation,
the dipole matrix element is represented by the left-most diagram
which corresponds to the photon absorption and electron emission from
the same shell. The inter-shell correlation allows for the photon
absorption by the shell $n_jl_j$ and the electron emission from the
shell $n_il_i$. This process is shown by the two remaining diagrams in
which the inter-shell correlation precedes or follows the photon
absorption (time-reverse and time-forward diagrams, respectively).
The exchange leads to the two ladder-type diagrams (not shown) in
addition to the two bubble-type diagrams shown in \Fref{Feynman}.

The same matrix element is used to evaluate the partial photoionization
cross-section from an occupied state $n_il_i$ to the photoelectron
continuum state $El$
\be
\label{CS} 
\sigma_{n_il_i\to El}(\omega) =
\frac43 \pi^2\alpha a_0^2\omega
\left| \la E l\,\|D\|n_il_i \ra \right|^2 \ .
\ee
Here $\alpha$ is the fine structure constant and $a_0$ is the Bohr
radius. The occupied orbitals $n_il_i$ and continuous orbitals $El$
are obtained by the self-consistent field and frozen-core field
Hartree-Fock methods, respectively, as in Ref \cite{AC97}. For the
present case of the $4d$ shell in Xe, correlations with the $5s$ and
$5p$ shells are included resulting in 5 non-relativistic channels. The
relativistic RPA (RRPA) method, is exactly the same as the
nonrelativistic version, except that it is based on the fully
relativistic Dirac equation, rather than the Schr\"{o}dinger equation,
and Dirac-Fock \cite{DF} instead of Hartree- Fock orbitals are
employed.  In the present RRPA calculation, final state correlations
are included through interchannel coupling between all the 13
relativistic dipole channels originating in the $5p$, $5s$ and $4d$
subshells which are open for photoionization in the energy range
studied in the present work; in other words with the relativistic
splittings of the five nonrelatistic channels become thirteen
channels. Omitting the inner shells has no significant effect upon the
results.  In both cases, the fullerene cage is modeled by a
square-well potential \eref{eqVc} with the following parameters:
$R_{\rm inner}=5.8$~a.u., $\Delta=1.9$~a.u., and $U_0=0.302$~au.

\begin{widetext}  
\begin{figure}[h]
\bp{12cm}
\vs{3.cm}
\hs{-7cm}
\epsfxsize=6cm
\epsffile{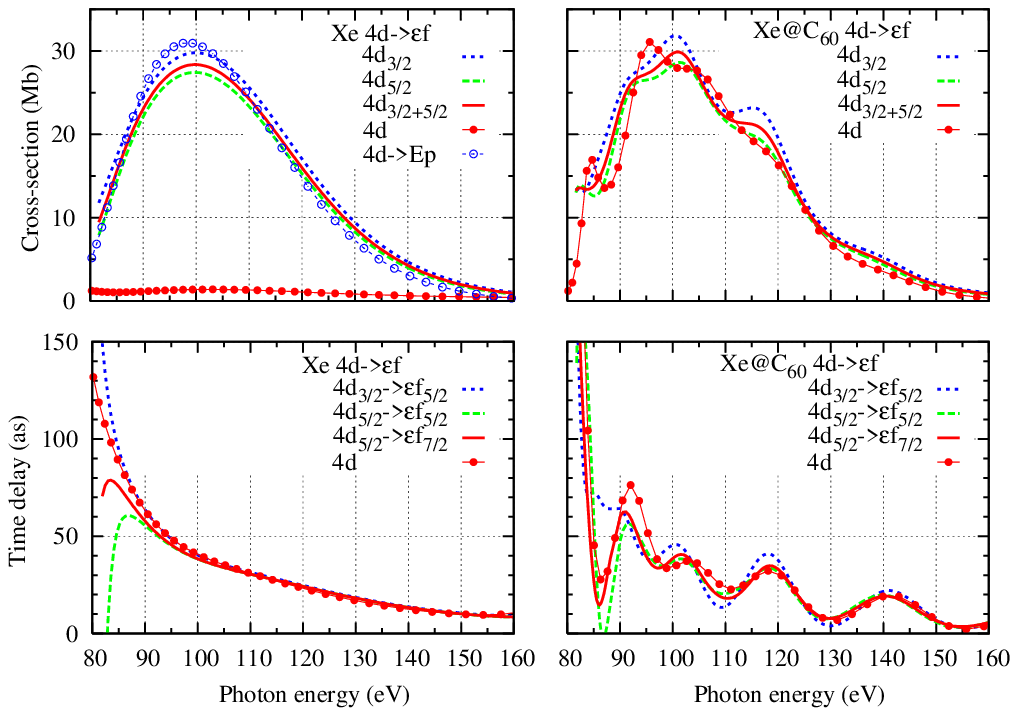}  
\ep
\bp{6cm}
\vs{3.cm}
\hs{-1cm}
\epsfxsize=6cm
\epsffile{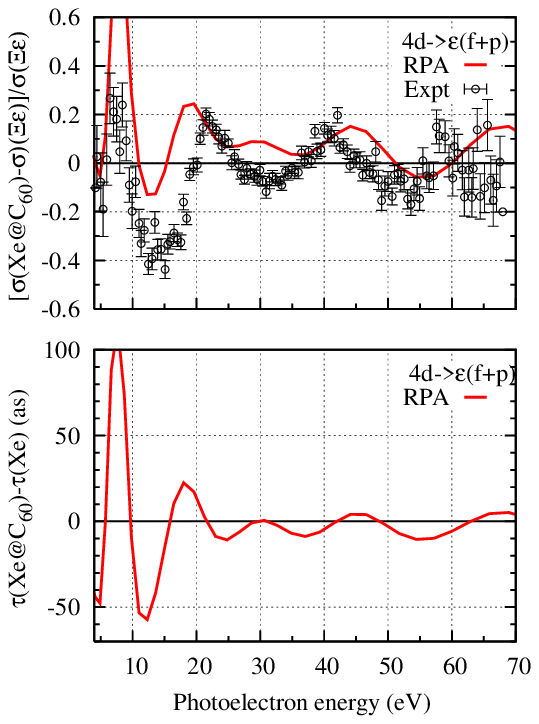}  
\ep
\vs{.5cm}
\bp{10cm}
\caption{ Top: partial photoionization cross-sections of the $4d$
shell in Xe (left) and Xe@\cx (right). The RRPA calculations for
spin-orbit resolved components $4d_{3/2}$ ($\times5/2$), $4d_{5/2}$
($\times5/3$) and their sum are drawn with lines. The RPA calculations
for the $4d\to \epsilon f$ and $4d\to\epsilon p$ (left panel only)
cross-sections are drawn with line-dots. Bottom: time delay in
photoionization of the $4d\to \epsilon f$ channel of Xe (left) and Xe@\cx
(right).  The RRPA calculations for various spin-orbit resolved
components are drawn with lines while the RPA calculation is drawn
with line-dots
\label{Data}
.}
\ep
\hs{5mm}
\bp{8cm}
\caption{Normalized  photoionization cross-section difference 
$\rm
\left[
\sigma(Xe@C_{60})-\sigma(Xe)
\right]
/\sigma(Xe)
$
(top) and time delay difference $\rm \tau(Xe@C_{60})-\tau(Xe)$ (bottom)
as functions of photoelectron energy. The RPA calculation is shown
with a solid line. The experimental data in the top panel
are from \cite{PhysRevA.88.053402}.
\vs{1.cm}
\label{Data1}
}
\ep
\end{figure}
\end{widetext}

The results of the calculations are shown in \Fref{Data}. In the upper
panels, the $4d$ partial photoionization cross-sections for
the free (left) and confined (right) Xe atoms are displayed. The RRPA results are
shown for the spin-resolved $4d_{1/2}$ and $4d_{3/2}$ subshells
leading to the $\epsilon f$ ionization continuum.  The cross-sections
are weighted with the inverse statistical factors (5/2 and 5/3) to
facilitate the shape comparison. The RPA results are shown for the
$4d\to\epsilon f$ and $4d\to\epsilon p$ transitions; the former is
clearly dominant in the given energy region. This allows us to
concentrate on the cross-section and time delay analysis in the
dominant channel only. The RPA and RRPA cross-section results are very
close for the free Xe atom after a small photon energy adjustment is
made to accommodate different $4d$ ionization thresholds in RRPA
(theoretical) and RPA (experimental). The cross-sections for Xe@\cx
are qualitatively similar between the two methods, although the RPA
predicts somewhat sharper resonances at lower photon energy end. This
difference between RPA and RRPA occurs owing to the interchannel
coupling among the spin-orbit-split relativistic channels which tends
to dampen the confinement resonances a bit.  This is known as
spin-orbit-activated interchannel coupling
\cite{PhysRevLett.88.093002, PhysRevA.79.043401}.

In the lower panels, the Wigner time delay results for
free (left) and confined (right) Xe atoms are shown.  Agreement of the two
methods for the free Xe atom are good except for lower photon energy
end where one sees a strong deviation between the spin-resolved $4d$
states; again this is caused by the spin-orbit-activated interchannel coupling. 
The RPA delay is close to the RRPA $4d_{3/2}\to\epsilon
f_{5/2}$ delay. For Xe@\cx, the two methods show 
the same set of
confinement resonances. The precise shape of the lowest resonance is
somewhat dependent of the spin-orbit splitting and the interchannel coupling.

Of particular importance is the finding that the confinement resonances are
much more prominent in the time delay than in the cross-section, thus making 
time-delay studies a much more sensitive probes of confinement resonances.  
In addition, the maxima in the time delay are not at the same energies as 
the maxima in the cross-sections.  This is not completely surprising 
because the the cross-section is quadratic with the amplitude, and the time 
delay is calculated from the logarithmic derivative  of the
amplitude.

A more detailed comparison of the cross-sections and time delays in
free and encapsulated Xe atoms is shown in \Fref{Data1}; for simplicity,
only the nonrelativistic RPA result is shown since the RRPA results are 
qualitatively similar. In the upper
panel of this figure, the normalized cross-section difference
$\rm \left[ \sigma(Xe@C_{60})-\sigma(Xe) \right] /\sigma(Xe) $ is plotted and
compared with the experimental data of
\cite{PhysRevA.88.053402}. Our RPA calculation qualitatively reproduces
all the resonances seen experimentally. A small offset
in peak position of the resonances may be attributed to the rough
spherical-well representation of the fullerene cage in the present
work.  In the bottom panel, an analogous plot for the time
delay difference $\rm \tau(Xe@C_{60})-\tau(Xe)$ is displayed. In this difference
representation, the oscillations on both plots are perfectly aligned, i.e.' 
the maima and minima in cross-section space and time-delay space occur 
at exactly the same energies.

Note also that, for both free and confined Xe, the three relativistic
$4d\to\epsilon f$ channels exhibit rather different time delays for
energies near the thresholds.  This is an indication that there are
real dynamical differences among the channels brought about by
interchannel coupling.  With increasing energy, these differences are
seen to disappear, and the time delays for all three relativistic
channels coalesce; not only do they coalesce with one another, but
they also become indistinguishable from the nonrelativistic result.
This occurs because the spin-orbit-activated interchannel coupling
decreases as the energy increases because the small difference in
thresholds becomes irrelevant as the photoelectron energy gets much
larger than the spin-orbit splitting \cite{PhysRevLett.88.093002,
PhysRevA.79.043401}.


\np

\np
\np In conclusion, in the present work, we have demonstrated clear
presence of confinement resonances in photoelectron group delay
(Wigner time delay) of the $4d$ shell of endohedrally confined Xe
atom. These resonances have been observed recently in the $4d$
photoionization cross-section of Xe@\cx
\cite{PhysRevLett.105.213001,PhysRevA.88.053402}. Our calculations
show that these resonances are even more prominent in the time
delay. This suggests that additional insights into the photoionization
process can be obtained by investigations performed in time-delay
space.  We further suggest that time delay experiment be used as an
effective novel way to study photoionization not only of neutral
endohedral $A$@C$_{60}$ but their charged members $A$@C$_{60}^{\pm z}$
as well as giant $A$@C$_{n >60}$ and multi-walled
$A$@C$_{60}$@C$_{240}$, $A$@C$_{60}$@C$_{240}$@C$_{540}$,
\textit{etc.}, endohedral fullerenes, each of which exhibits its own
specific photoionization properties \cite{Dolmatov2009}. Also of
interest are studies of time delays of other elementary processes
involving endohedral fullerenes, e.g., elastic electron scattering off
$A$@C$_{60}$, where initial insight has been provided recently
\cite{2014arXiv1401.2632D}.

We note that in the present work we did not consider the influence of
the probing IR field on the measured time delay. This effect, however,
was analyzed in \cite{Nagele2014} where it was found that the laser
Coulomb coupling correction and the screening effect of the \cx shell
nearly totally cancel each other and hence the measured time delay
will be essentially the Wigner time delay.

Finally, we urge experimentalists to initiate time delay studies of
 endohedral fullerenes for which the well established attosecond
 streaking \cite{M.Schultze06252010} and RABITT
 \cite{PhysRevLett.106.143002} techniques could prove to be useful.

%
This work was supported by the Department of Science and
Technology (DST), Government of India, by the National Science 
Foundation, by the Department of Energy, 
Office of Chemical Sciences, and by the Australian Research 
Council.

\np

\end{document}